\documentclass[%
mathleft,%
final,%
]{an}
\usepackage{graphicx}
\usepackage{times}
\begin{document}

\Pagespan{1}{}
\Yearpublication{2014}%
\Yearsubmission{2013}%
\Month{1}%
\Volume{335}%
\Issue{1}%
\DOI{This.is/not.aDOI}%

\title{The chemical composition of the Small Magellanic Cloud \thanks{based on observations obtained at Paranal ESO
Observatory with the FLAMES facility.}}

\author{A. Mucciarelli\inst{1}\fnmsep\thanks{Corresponding author.
  \email{alessio.mucciarelli2@unibo.it}}
}
\titlerunning{SMC chemical composition}
\authorrunning{A. Mucciarelli}
\institute{Dipartimento di Fisica e Astronomia, Universit\`a degli Studi
di Bologna, v.le Berti Pichat 6/2, I$-$40127 Bologna, Italy}

\received{XXXX}
\accepted{XXXX}
\publonline{XXXX}

\keywords{Magellanic Clouds, stars: abundances, techniques: spectroscopic}

\abstract{The Small Magellanic Cloud is a close, irregular galaxy 
that has experienced a complex star formation history due to the 
strong interactions occurred both with the Large Magellanic Cloud and the 
Galaxy.
Despite its importance, the chemical composition of its stellar 
populations older than $\sim$1-2 Gyr is still poorly investigated.
I present the first results of a spectroscopic survey of $\sim$200 
Small Magellanic Cloud giant stars performed with FLAMES@VLT. 
The derived metallicity distribution peaks at [Fe/H]$\sim$--0.9/--1.0 dex, 
with a secondary peak at [Fe/H]$\sim$--0.6 dex. All these stars show 
[$\alpha$/Fe] abundance ratios that are solar or mildly enhanced ($\sim$+0.1 dex). Also, three 
metal-poor stars (with [Fe/H]$\sim$--2.5 dex and enhanced [$\alpha$/Fe] ratios 
compatible with those of the Galactic Halo) have been detected in the outskirts of the 
SMC: these giants are the most metal-poor stars discovered so far in the Magellanic Clouds.}

\maketitle

\section{Introduction}
In the last decade, the advent of the high-resolution spectrographs 
mounted on 8-10 m class telescopes (i.e. VLT, Keck, Subaru) has allowed us 
to investigate the chemical composition of individual Red Giant Branch (RGB)
stars in the Local Group, extending our knowledge in galactic environments outside our Galaxy. 
Detailed chemical abundances analysis for old ($>$1-2 Gyr) stellar 
populations are available for the Sagittarius remnant (\cite{sbordone07}), 
for the Large Magellanic Cloud (LMC, \cite{pompeia,m08,m10,m11,m12}) and for 
some isolated nearby dwarf galaxies (\cite{shetrone01,shetrone03}, \cite{letarte10}, \cite{lemasle12}).
Despite its proximity ($\sim$60 kpc), the chemical composition of the Small 
Magellanic Cloud (SMC) is still poorly known, without available studies 
based on high-resolution spectroscopy concerning its old stellar populations.
This irregular galaxy is characterized by a relevant and ongoing star-formation 
activity. The SMC has experienced a complex and violent star formation history, 
because it is gravitationally bound with the LMC and the Milky Way, forming a 
triple system. The mutual tidal interactions occurring among these three galaxies 
have probably triggered the main star formation episodes in the Magellanic Clouds 
(see for instance \cite{bekki05}).
Thus, the study of the chemical composition of the stars in the SMC is a gold-mine of 
information to understand the chemical enrichment history of irregular galaxies 
characterized by tidal interactions and matter exchanges.

Currently, our knowledge of the chemical composition of the stars in the SMC 
is based on high-resolution spectroscopy of supergiant stars 
(\cite{spite89a}, \cite{spite89b}, \cite{hill97}, 
\cite{luck98}, \cite{venn99}), thus sampling only the youngest ($<$200 Myr) 
stellar populations, and on low-resolution spectroscopy through the Ca~II triplet 
(\cite{carrera08}), providing metallicities for stellar populations older than $\sim$1 Gyr 
but without information about the individual elements.
In this contribution, I discuss the preliminary results of a spectroscopic survey of the SMC giant stars performed 
with the ESO spectrograph FLAMES with the final aim to provide a global and deep 
comprehension of the chemical enrichment history of this galaxy.

\section{Observations}
High-resolution spectra have been acquired with the multi-object 
facilities FLAMES (\cite{pasquini}) mounted at the UT2 of the 
Very Large Telescope during 4 nights of observations. 
Three SMC fields have been selected around three different globular clusters, 
thus to observe both cluster stars (with the UVES fibers, Lapenna et al., in prep.) 
and the surrounding field stars (with the GIRAFFE fibers, discussed here, Mucciarelli et al., in prep.). 
The three targets globular clusters have been selected in order to sample different 
regions of the SMC and different ages: {\sl (i)} NGC~419, an intermediate-age cluster 
($\sim$1-2 Gyr) in the innermost region of the SMC, {\sl (ii)} NGC~339, an external 
cluster with an age of $\sim$5-6 Gyr, and {\sl (iii)} NGC~121, the unique old ($\sim$11 Gyr) 
SMC globular cluster, located in the outskirts of the SMC.

Two GIRAFFE gratings (namely HR11 and HR13) have been secured for all the 
targets, allowing to measure Fe and $\alpha$-elements (O, Mg, Si, Ca and Ti) and 
other key-elements (like Na, Ni and Ba).
The targets have been selected in the bright portion of the RGB, by using 
the near-infrared SOFI@NTT photometry (\cite{msmc}) for the regions close 
to the cluster (within a radius of $\sim$2.5 arcmin from the cluster center) 
and the 2MASS photometry for the outermost regions. 

The dataset includes a total of 214 giant stars, members of the SMC field according to their 
radial velocities; this is the largest dataset of high-resolution spectra for SMC field stars 
collected so far. 

\section{Analysis}
Atmospherical parameters have been derived from the photometry, combining the near-infrared 
SOFI+2MASS colors with the optical colors provided by the Magellanic Photometric Survey 
(\cite{zar}). The traditional method of the equivalent widths (EWs) has been adopted to infer 
the chemical abundances of the investigated elements. EWs of unblended lines have been measured with the automatic 
code DAOSPEC (\cite{stetson}) and the abundances derived by using the package GALA\footnote{http://www.cosmic-lab.eu/gala/gala.php} 
(\cite{mgala}), coupled with the last generation of ATLAS~9 model atmospheres (\cite{cas}).
Spectral synthesis (through a 
$\chi^2$-minimization between the observed spectrum and a grid of suitable synthetic spectra computed 
with the code SYNTHE) has been 
used only for O (to take into account the blending of the forbidden O~I line at 6300.3 $\mathring{A}$ 
with a close Ni~I transition).


\section{Results}
\subsection{The metallicity distribution}
\begin{figure}
\includegraphics[angle=0, width=\linewidth]{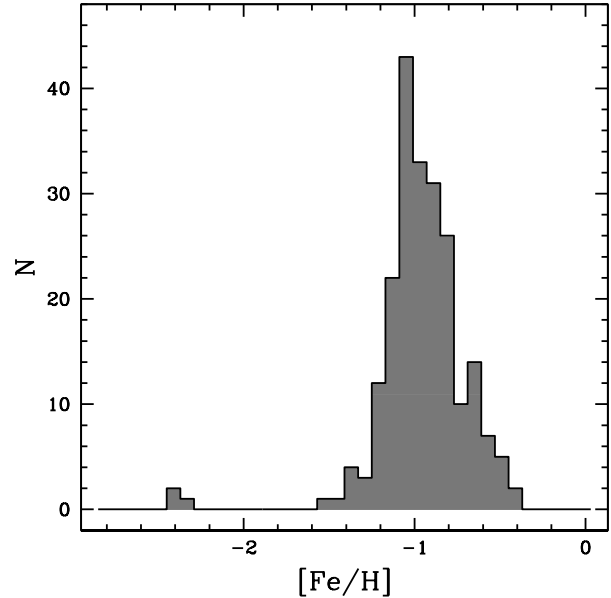}
\caption{The metallicity distribution of the entire sample of SMC 
giant stars observed with FLAMES-GIRAFFE.}
\label{md}
\end{figure}

\begin{figure}
\includegraphics[angle=0, width=\linewidth]{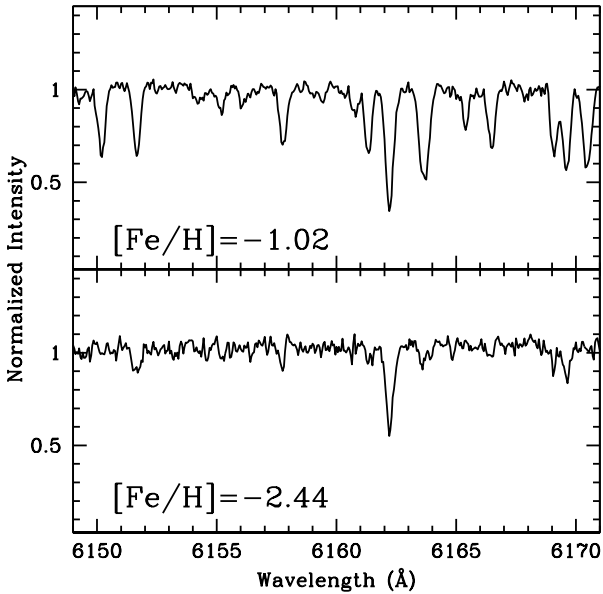}
\caption{Comparison between the spectra of two SMC giant stars 
observed with the FLAMES-GIRAFFE HR13 setup, with similar atmospheric 
parameters and different metallicity, namely [Fe/H]=--1.02 dex (upper panel) 
and [Fe/H]=--2.44 dex (lower panel).}
\label{spec}
\end{figure}

Fig.~\ref{md} shows the metallicity distribution of the entire 
sample, ranging from [Fe/H]$\sim$--2.6 dex to [Fe/H]$\sim$--0.4 dex.
We can distinguish three main features in the SMC metallicity distribution:\\ 
(1)~a mean peak
at [Fe/H]$\sim$--0.9/--1.0 dex, that we can consider as the 
typical metallicity of the old stellar populations in the SMC. 
This metallicity is consistent with those observed in the old 
globular clusters NGC~121 and NGC~339, suggesting that these stars formed 
during the first $\sim$5 Gyr of the evolution of the SMC. This scenario 
seems to be in agreement with the star formation history of the SMC 
derived by \cite{harris}, where approximately 50\% of the SMC stars formed 
in early epochs (at ages $>$8 Gyr ago);\\ 
(2)~a secondary peak at [Fe/H]$\sim$--0.6 dex, probably linked 
to the star formation burst occurring in the last few Gyr and due 
to the tidal capture of the SMC by the LMC (\cite{bekki05}). 
This burst is also responsible of the 
formation of some intermediate-age SMC globular clusters
(like NGC~419) and of the metal-rich ([Fe/H]=~--0.5 dex) population 
observed in the metallicity distribution of the LMC (\cite{lapenna});\\ 
(3)~a metal-poor tail, including three giants in the field of NGC~121, 
with [Fe/H]$\sim$--2.5 dex.
This is the first discovery of metal-poor ([Fe/H]$<$--2.0 dex) 
field stars in the Magellanic Clouds (in fact, the most metal-poor 
field star known so far in the Magellanic Clouds is a LMC giant with [Fe/H]=~-1.74 dex, 
see \cite{pompeia}).
Fig.~\ref{spec} shows the comparison between the spectra taken with the 
HR13 setup of one of these metal-poor stars and a giant with 
the typical metallicity of the SMC.\\
Our survey confirms, for the first time through the analysis of a large sample 
of high-resolution spectra, 
that the bulk of the old stellar populations in the SMC has a metallicity lower 
than that observed in the LMC, where the dominant population has [Fe/H]$\sim$--0.5 dex 
(\cite{lapenna}). The derived metallicity distribution nicely matches with the scenario 
proposed by \cite{harris} for the chemical enrichment history of the SMC: 
the bulk of the stars formed in the first bursts of star formation or in the 
following long quiescent period with a low star formation efficiency (until about 
3 Gyr ago) should show low-metallicity ([Fe/H]$\sim$--1 dex), with a following increase 
of the metallicity (reaching [Fe/H]$\sim$--0.4 dex) due to the recent bursts of 
star formation (and linked to the past close encounter of the SMC with the LMC 
and the Milky Way). 

\subsection{$\alpha$-elements}
\begin{figure}
\includegraphics[angle=0, width=\linewidth]{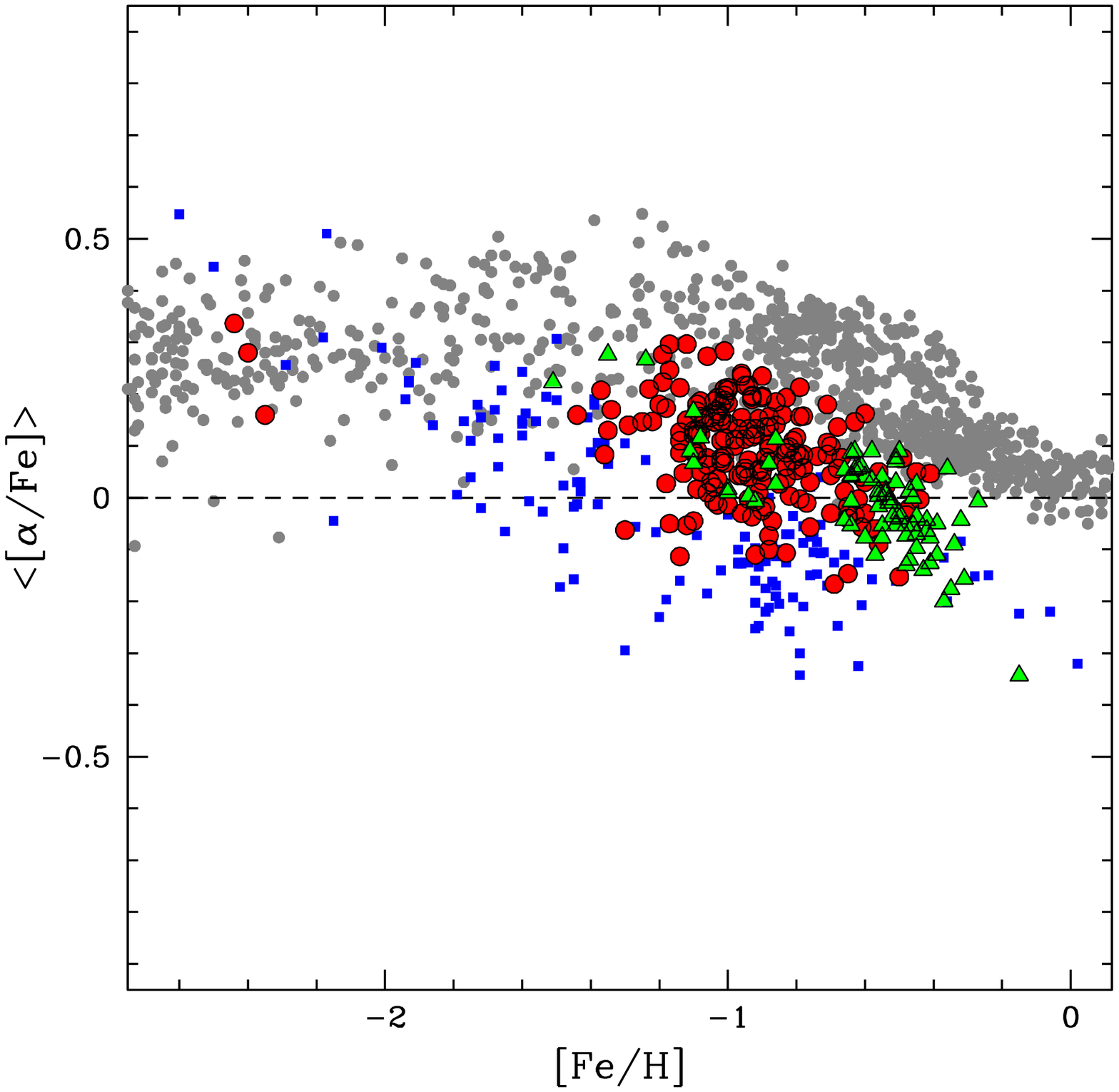}
\caption{Behavior of the average [$\alpha$/Fe] abundance ratio 
(obtained by averaging O, Mg, Si, Ca and Ti abundances) 
as a function of [Fe/H] for the SMC giant stars observed 
with FLAMES-GIRAFFE (red circles), in comparison 
with the abundance ratios obtained for stars in the Milky Way stars 
(small grey circles, \cite{edvardsson93, gratton03, reddy03, barklem05, reddy06}), 
in the LMC (green triangles, \cite{lapenna}) 
and dwarf spheroidal galaxies of the Local Group 
(blue squares, \cite{shetrone01, shetrone03, sbordone07, letarte10, lemasle12}).
}
\label{alf}
\end{figure}

The [$\alpha$/Fe] ratio represents a powerful diagnostic to investigate 
the relative role played by Type II Supernovae (main producers of $\alpha$-elements) 
and Type Ia Supernovae (main producers of Fe) in the chemical enrichment 
process. Indeed, there is time delay between the explosion of Type II Supernovae, 
occurring since the onset of the star formation event, and Type Ia Supernovae, which 
happen later on (\cite{greggio05}).\\ 
Fig.~\ref{alf} shows the behavior of the $<$[$\alpha$/Fe]$>$ abundance ratio 
as a function of [Fe/H] for the SMC field stars (red points), in comparison 
with stars analyzed with high-resolution spectroscopy (Milky Way, grey points; 
LMC, green triangles, dwarf spheroidal galaxies, blue squares). 
Typically, the $<$[$\alpha$/Fe]$>$ abundance ratios in the SMC giants 
turn out to be roughly solar (or mildly enhanced, $\sim$+0.1 dex). 
The overall trend of $<$[$\alpha$/Fe]$>$ 
ratio with the iron content shows a decrease at increasing metallicity. 
Also, the $<$[$\alpha$/Fe]$>$ ratio
is systematically lower than those observed in the Milky Way stars of similar metallicity, but 
significantly higher (of about 0.2-0.3 dex) than those measured in the dwarf spheroidal galaxies. 
The three metal-poor stars detected in the field of NGC~121 show a general agreement 
with the $<$[$\alpha$/Fe]$>$ ratios observed in the Galactic Halo stars of similar metallicity.

These findings suggest that the bulk of the SMC stars formed from a gas 
enriched by the ejecta of both Type Ia and Type II Supernovae, in agreement 
with the expectations that environments characterized by star formation rates 
less efficient than that of the Milky Way will show the so-called {\sl knee} in the 
[$\alpha$/Fe]--[Fe/H] plane at metallicity lower than [Fe/H]$\sim$--1 dex.
It is worth noticing that the stars in LMC and SMC seem to define an unique sequence in 
the [$\alpha$/Fe]--[Fe/H] plane, clearly distinct with respect to that defined by the 
dwarf spheroidal galaxies and that of the Milky Way.
Interesting enough, the giant stars of Sagittarius remnant well agree with the 
mean locus defined by the Magellanic Clouds stars, suggesting a different chemical 
enrichment history between the isolated, dwarf galaxies (with lower $<$[$\alpha$/Fe]$>$ ratios) 
and the galactic environments characterized by strong interactions with other systems 
(like LMC, SMC and Sagittarius).

\section{Conclusions}
In this contribution, I have discussed the first results (about the metallicity 
distribution and the $\alpha$-elements abundances) of a spectroscopic campaign 
performed with FLAMES@VLT to investigate the chemistry of the SMC giant stars. 
These results allow for the first time to have a global picture of the chemical 
composition of the SMC stellar populations older than $\sim$1-2 Gyr, providing 
a valuable tool to understand the chemical enrichment history of this galaxy, 
in particular in light of the mutual interactions among the LMC, SMC and the Milky Way.
Finally, the discovery of the most metal-poor giant stars observed so far in the Magellanic 
Clouds opens a new perspective in the study of these irregular galaxies, allowing 
to shed new light on the early epochs of star formation of the SMC.


%
%

\end{document}